\documentclass[aps,12pt, pra,showpacs,showkeys,nofootinbib,superscriptaddress,longbibliography]{revtex4-1}

\usepackage{graphicx}
\usepackage{enumerate}
\usepackage{amsfonts}
\usepackage[a4paper]{geometry}
\usepackage{amsmath}
\usepackage{amssymb}
\usepackage{amsbsy}     
\usepackage{latexsym}   
\usepackage{setspace} 
\usepackage{fancyhdr}
\usepackage{epsfig}
\usepackage[usenames]{color}
\usepackage{caption}
\usepackage{bm}
\usepackage{float}

\definecolor{dred}{rgb}{.8,0.2,.2}
\definecolor{ddred}{rgb}{.8,0.5,.5}
\definecolor{dblue}{rgb}{.2,0.2,.8}
 
\newcommand{\eye}{\mathbf{1}}

\newtheorem{theorem}{Theorem}[section]

\newtheorem{lemma}[theorem]{Lemma}

\newtheorem{definition}[theorem]{Definition}

\newtheorem{remark}[theorem]{Remark}
\newtheorem{result}[theorem]{Result}

\newcommand{\ket}[1]{\left|#1\right\rangle}
\newcommand{\bra}[1]{\left\langle #1 \right|}
\newcommand{\braket}[2]{\left\langle #1 \hspace{1mm} \vline \hspace{1mm} #2 \right\rangle}

\newcommand{\uket}[1]{|#1\rangle}
\newcommand{\ubra}[1]{\langle #1 |}

\def\1#1{{\bf #1}}
\def\2#1{{\cal #1}}
\def\3#1{{\sl #1}}
\def\4#1{{\tt #1}}
\def\5#1{{\sf #1}}
\def\6#1{{\mathfrak #1}}
\def\7#1{{\mathbb #1}}

\newcommand{\be}{\begin{equation}}
\newcommand{\ee}{\end{equation}}

\newcommand{\op}[1]{\bm{#1}}
\newcommand{\spc}[1]{\mathcal{#1}}
\newcommand{\grp}[1]{\mathrm{#1}}

\DeclareMathOperator{\Tr}{Tr}

\definecolor{DarkRed}{rgb}{0.7,0,0}

\newcounter{ComntCntr} 

\newcommand{\captionfonts}{\small}

\makeatletter  
\long\def\@makecaption#1#2{%
  \vskip\abovecaptionskip
  \sbox\@tempboxa{{\captionfonts #1: #2}}%
  \ifdim \wd\@tempboxa >\hsize
    {\captionfonts #1: #2\par}
  \else
    \hbox to\hsize{\hfil\box\@tempboxa\hfil}%
  \fi
  \vskip\belowcaptionskip}
\makeatother   
\graphicspath{{pics/}}

\begin{document}
\title{Tensor Networks for Entanglement Evolution}

\author{Sebastian Meznaric}
\affiliation{Clarendon Laboratory, University of Oxford}

\author{Jacob Biamonte}
\email{jacob.biamonte@qubit.org}
\affiliation{Institute for Scientific Interchange, Torino Italy}
\affiliation{Centre for Quantum Technologies, National University of Singapore}



\begin{abstract}
The intuitiveness of the tensor network graphical language is becoming well known through its use in numerical simulations using methods from tensor network algorithms. Recent times have also seen rapid progress in developing equations of motion to predict the time evolution of quantum entanglement [Nature Physics, 4(\textbf{4}):99, 2008].  Here we cast these recent results into a tensor network framework and in doing so, construct a theory which exposes the topological equivalence of the evolution of a family of entanglement monotones in arbitrary dimensions.  This unification was accomplished by tailoring a form of channel state duality through the interpretation of graphical tensor network rewrite rules.  
The introduction of tensor network methods to the theory of entanglement evolution opens the door to apply methods from the rapidly evolving area known as tensor network states.   
\end{abstract}
\maketitle

\section{Introduction}

There has been increasing recent interest in the topic of evolution of quantum entanglement --- inspired by the work \cite{Konrad08}. Konrad et al. considered the evolution equation for quantum entanglement for qubits \cite{Konrad08} and its later extension to higher dimensions can be found in \cite{Tiersch09, Li09-1}. The multipartite case was characterized by Gour \cite{Gour10}. Before \cite{Konrad08}, people have generally dealt with entanglement evolution on a case-by-case basis. In this spirit, several authors considered non-Markovian entanglement evolution \cite{Bellomo07,  Yuan07, Dajka08, Yu10}. Similarly, upper and lower bounds can be obtained for special classes of states such as graph states \cite{Cavalcanti09}. 

Our approach aims to develop a tensor network diagrammatic method and to show which aspects of entanglement evolution can be described in those terms. We found that key results of Tiersch \cite{Tiersch09} can be reproduced in the language of tensor networks with several appealing features. Konrad et al. themselves developed and relied on a sequence of diagrams, that aided in their exposition of the topic at hand \cite{Konrad08}.  These diagrams were utilitarian in nature and used to convey a point.  However, there exists a diagrammatic graphical language, of which Hilbert spaces are a complete model, known as Penrose tensor networks or string diagrams \cite{Penrose}.  These diagrams can aid in intuition and also represent mathematical equations \cite{Baez09}.

The diagrammatic language built on to describe the tensor networks dates back to Penrose \cite{Penrose}.  Recently, we have also done work on the theory and expressiveness of tensor network states \cite{CTNS, BB11, sam, wood11}.  What we add to the literature on tensor network states is the development of some theory which connects these ideas to a specific version of channel-state duality.  This offered an advancement and extension of the prior graphical aid of \cite{Konrad08}.  Through this rigorous graphical language a unified common topological structure of the networks representing the evolution of entanglement also emerges.  

The introduction of a tensor network theory of entanglement evolution opens the door to apply numerical methods from the rapidly evolving area known as tensor network states \cite{MPSreview08, TNSreview09, tomi}.  In recent times, tensor networks states have been successfully used to simulate a wide class of quantum systems using classical computer algorithms.  In addition, theory has been developed to use these methods to simulate classical stochastic systems \cite{tomi}. At the heart of these methods are tensor network contraction algorithms, pairing physical insight with a new theory behind making approximations to truncate and hence minimize the classical data needed to represent a quantum state.  The key properties and functions of these algorithms are known to be described in the graphical language we have adapted to entanglement evolution \cite{Penrose, CTNS, BB11}.  

Structure.  We will begin in Section \ref{sec:notation} by reviewing those tensor network components that we will use.  This introduction paves the way for a review of the so called snake equation.  This equation is then used to prove a graphical version of map-state duality --- inspired by and tailored specifically to the problem at hand.  Section \ref{sec:evolution} then applies these building blocks and associated tools from tensor network states to the theory of entanglement evolution.  We begin by recalling the fundamental definitions and a quick outline of the known theory, we then cast these methods into our framework and pinpoint a wide class of measures which all share the same network topology in the tensor network language.  The conclusion presents an overview of potential future directions by listing some open problems and new directions.

\section{Penrose graphical notation and map-state duality}\label{sec:notation}

We begin by showing how quantum states and operators may be represented graphically using the tensor network language. We then proceed to more complex topics, including the state-map duality and its graphical equivalents. 

It is typical in quantum physics to fix the computational basis and consider a tensor as an indexed multi-array of numbers. For instance,  
\begin{center}
 \includegraphics[width=6cm]{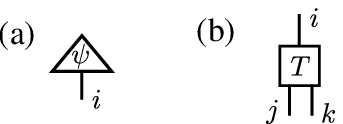}
\end{center}
represent the tensor (a) $\psi_i$ and (b) $T^i_{~jk}$. The scenario readily extends.  

Graphically a two party quantum state $\ket{\psi}$ would have one wire for each index.  We can write $\ket{\psi}$ in abstract index notation as $\psi_{ij}$ or in terms of the Dirac notation convention (which we mainly adopt here) as $\ket{\psi} = \sum_{ij} c_{ij}\ket{i}\ket{j}$.  Appropriately joining a flipped (transpose) and conjugated (star or overbar) copy of the quantum state $\ket{\psi}$ allows one to represent the density operator $\rho = \sum c_{ij}\overline{c}^{kl}\ket{ij}\bra{kl}$ as follows \cite{Penrose}. 
\begin{figure}[H] 
\begin{center}
 \includegraphics[width=6cm]{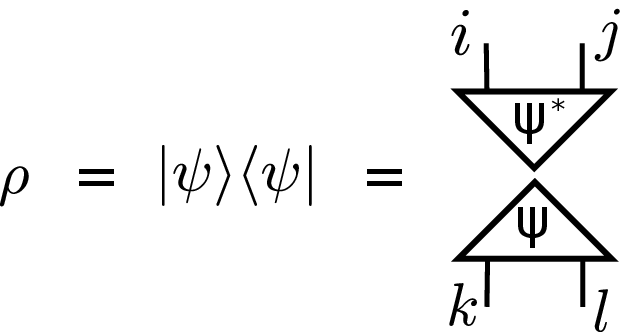}
\end{center}
\caption{Penrose's graphical representation of a pure density matrix.}
\end{figure}
Here a box with two inputs ($i$, $j$) and two outputs ($k$, $l$) would be used to represent a general density operator $\rho = \sum \rho_{ij}^{~~kl}\ket{ij}\bra{kl}$, 
with no known additional structure. Those familiar with quantum circuits will recognize this graphical language as their generalization \cite{BB11}.   

\begin{remark}[Tensor valence]
A tensor with $n$ indices up and $m$ down is called a valence-$(n,m)$ tensor and sometimes a valence-$k$ tensor for $k=n+m$.  
\end{remark}

\begin{remark}[Diagram convention --- top to bottom, or right to left]
Open wires pointing towards the top of the page, correspond to upper indices (bras), open wires pointing towards the bottom of a page correspond to lower indices (kets).  For ease of presentation, we will often rotate this convention $90$ degrees counterclockwise.
\end{remark}

We are interested in the following operations: (i) tensor index contraction by connecting legs of different tensors, (ii) raising and lowering indices by bending a leg upwards or downwards, respectively and thus taking the appropriate conjugate-transpose and (iii) a duality between maps, states and linear maps in general, called \textit{Penrose wire bending duality}. This duality will play a major role in our application of the language to entanglement evolution. The duality is obtained by noticing that states are tensors with all open wires pointing in the same direction.  Bending some of the wires in the opposite direction makes various linear maps out of a given state. Given the computational basis, this amounts to turning all kets belonging to one of the Hilbert spaces into bras and vice versa. A bipartite state in the fixed standard basis 
\begin{align}
	\ket{\psi} = \sum_{i,j} A_{i,j}\ket{i}\ket{j},
\end{align}
for example, is dual to the operator
\begin{align}
	[\psi] = \sum_{i,j} A_{i,j} \ket{i}\bra{j}.
\end{align} 
This result can be shown by considering components of the equation $(A\otimes \eye)\ket{\Phi^+}$ 
where $\ket{\Phi^+}$ is the Bell state \eqref{eqn:delta-bell}.  

The Choi-Jamio{\l}kowski isomorphism can be cast as a case of Penrose Duality.  To do this we must recall 
Penrose's cups, caps and identity wires.  As in \cite{Penrose}, these three tensors are given diagrammatically as 
\begin{center}
 \includegraphics[width=12cm]{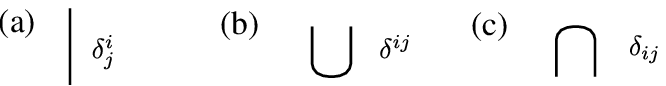}
\end{center}
By thinking of these tensors now in terms of components, e.g.\ $\delta_{ij}$ is $1$ whenever $i=j$ and $0$ otherwise, we note that 
\begin{equation}\label{eqn:delta}
 \eye = \sum_{ij}\delta^i_j\ket{i}\bra{j}=\sum_k\ket{k}\bra{k} 
\end{equation}
\begin{equation}\label{eqn:delta-effect}
 \bra{00}+\bra{11} + \cdots + \bra{nn} = \sum_{ij} \delta^{ij}\bra{ij}=\sum_k\bra{kk}
\end{equation}
\begin{equation}\label{eqn:delta-bell}
 \ket{00}+\ket{11} + \cdots + \ket{nn} = \sum_{ij} \delta_{ij}\ket{ij}=\sum_k\ket{kk}
\end{equation}
where the identity map (a) corresponds to Equation \eqref{eqn:delta}, the cup (b) to \eqref{eqn:delta-effect} and the cap (c) to \eqref{eqn:delta-bell}.  
The relation between these three equations is again given by Penrose's wire bending duality: in 
a basis, bending a wire corresponds to changing a bra to a ket, and vise versa, allowing one to 
translate between Equations \eqref{eqn:delta}, \eqref{eqn:delta-effect} and \eqref{eqn:delta-bell} at will.  

The contraction of two tensor indices diagrammatically amounts to appropriately joining open wires.  Given tensors 
$T^i_{~jk}$, $A^l_{~n}$ and $B^m_{~q}$ we form a contraction by multiplying by $\delta^j_l\delta^q_k$ resulting in the tensor
\be 
T^i_{~jk}A^j_{~l}B^k_{~m} := \Gamma^i_{~lm} 
\ee 
where we use the Einstein summation convention (repeated indices are summed over) and the tensor $\Gamma^i_{~lm}$ is introduced per definition 
to simplify notation.  In quantum physics notation, this is typically expressed in equational form as 
\be 
\Gamma = \sum_{ilm}\Gamma^i_{~lm}\ket{lm}\bra{i} = \sum_{ijklm} T^i_{~jk}A^j_{~l}B^k_{~m}\ket{lm}\bra{i}. 
\ee

\begin{result}[Graphical trace --- Penrose \cite{Penrose}]
Graphically the trace is performed by appropriately joining wires.  The following depiction in Figure \ref{fig:penrose-trace} (b) is Penrose's representation of a reduced density operator, where the stars on the $\Psi$'s represents complex conjugation. 
\begin{figure}[h]
\begin{center}
 \includegraphics[width=8cm]{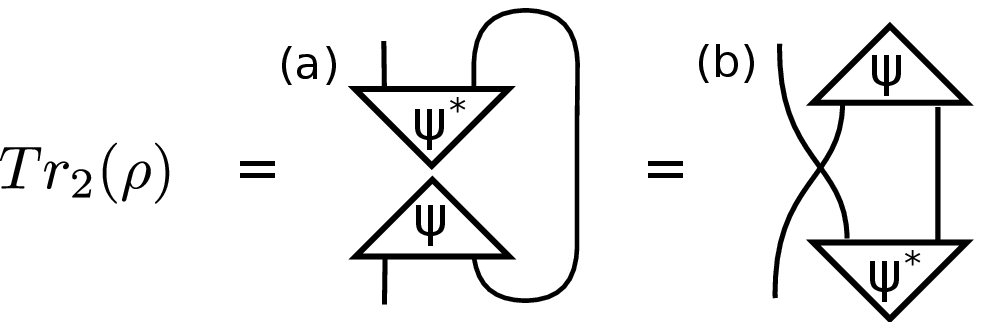}
\end{center}
\caption{Graphical representation of a partial trace. The network (b) is due to Penrose. (Note that Penrose used reflection across the page to represent adjoint, yet we have placed a star on $\Psi$ to represent conjugate for illustrative purposes.)\label{fig:penrose-trace}}
\end{figure}
\end{result}

We have now presented the key tensor network building blocks used in this study.  In practice, tensor networks contain an increasing number of tensors, 
making it difficult to form expressions using (inherently one-dimensional) equations.  The two-dimensional diagrammatic depiction of tensor networks can simplify such expressions and 
often reduce calculations.  Here we tailor this graphical tool to depict internal structure and lend insight into 
presenting a unified theory of entanglement evolution.  A key component of this unified view relies 
on the \textit{natural equivalence} induced by the so called \textit{snake equations}, which we will review next.  

The snake equation \cite{Penrose} amounts to considering the transformation of raising or lowering an index.  One can raise an index and then lower this index or vice versa, which amounts essentially to the net effect of doing nothing at all.  This is captured diagrammatically by the so called, \textit{snake} or \textit{zig-zag equation}, as 
\begin{center}
 \includegraphics[width=8cm]{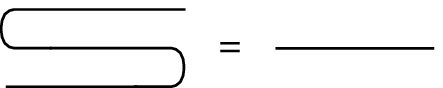}
\end{center}
together with its vertical reflection across the page.  In tensor index notation, it is given by $\delta_{ji}\delta^{ik} = \delta_j^{~k}$.

As a tool to aid in our analysis of entanglement evolution, we have noticed a succinct form of map state duality that works by considering 
the natural equivalence found by the snake equations inside an already connected diagram.  We expect that this formulation and its interpretation will have applications outside of the area of entanglement evolution.  To date, the snake equation has already been used to model quantum teleportation \cite{kauf05, 2009arXiv0910.2737A} and so the interpretation here is a different one. 

\begin{lemma}[Graphical map-state duality]\label{lemma:graphical-map-state}
Every bipartite quantum state $\ket{\psi}$ is naturally equivalent to a single sided map $[\psi]$ acting on the Bell state $\ket{\Phi^+}$ with graphical depiction as follows.  
\begin{figure}[h]
\begin{center}
 \includegraphics[width=12cm]{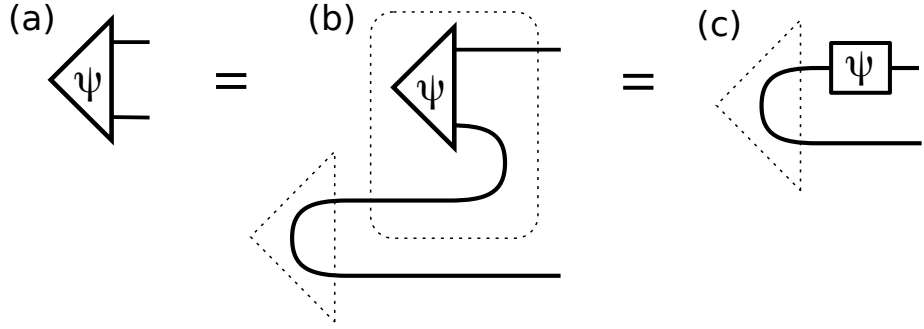}
\end{center}
\caption{Graphical channel state duality of a state acted on by the identity channel. (a) The state $\ket{\psi}$ can be thought of as being acted on by the trivial or identity channel on both sides.  (b) From the snake equation an identity wire can be equivalently replaced with its `S' shaped deformation.  This can then be thought of as a map $[\psi]$ (enclosed by a box with dashed lines on its edges) acting on a Bell state (enclosed with a triangle with dashed lines on its edges).  (c) An equivalent expression found by redrawing the shape of the tensor representing $\ket{\psi}$, into a box. \label{fig:map-state-duality-2}}
\end{figure}
\end{lemma}

We will make a few remarks related to Lemma \ref{lemma:graphical-map-state}. To prove the result using equations (instead of diagrams which was already done) we proceed as follows.  To consider $\ket{\psi}$ as a map $[\psi]$, we act on it with a Bell state as $[\psi]:=(\eye \otimes \bra{\Phi^+})\ket{\psi}$. (Notice that the net effect is to turn kets of one of the spaces into bras and vice versa, i.e. $\ket{0}\otimes \ket{0}$ becomes $\ket{0}\bra{0}$.) We then 
note that $\ket{\psi} = ([\psi]\otimes \eye) \ket{\Phi^+}$ and hence that $\ket{\psi} = ([(\eye \otimes \bra{\Phi^+})\ket{\psi}]\otimes \eye) \ket{\Phi^+}$.  The result seems more intuitive by considering the graphical rewrites.  

To explain the diagrammatic manipulations, begin with Figure \ref{fig:map-state-duality-2} (a), representing the state $\ket{\psi}$.  We can think of this (vacuously, it would seem) as a state being acted on by the identity operation.  Application of the snake-equation to one of the outgoing wires allows one to transform this into the equivalent diagram (b).  We can think of (b) as a Bell state (left) being acted on by a map found from coefficients of the state $\ket{\psi}$.  We have illustrated this map acting on the bell state in (b) by a light dashed line around $\ket{\psi}$. We then rewrite the figure, capturing the duality and arriving at (c).  

\begin{remark}[Properties of the map $\psi$]
 The map $[\psi]:=(\eye \otimes \bra{\Phi^+})\ket{\psi}$ evidently has close properties with the quantum state $\ket{\psi}$.  
 For instance, $\text{Tr}[\psi][\psi]^\dagger = \braket{\psi}{\psi}$.  The state $\ket{\psi}$ is invariant under the 
 permutation group on two elements iff $[\psi]=[\psi]^\top$.  The state $\ket{\psi}$ is SLOCC equivalent to the generalized Bell state
 iff $\det [\psi] \neq 0$ and the operator rank of $[\psi]$ gives the Schmidt rank of $\ket{\psi}$.  Rank is a 
 natural number and although it is invariant under action of the local unitary group, 
 it is not what is called a polynomial invariant or algebraic invariant since rank is not 
 expressed in terms of a polynomial in the coefficients of the state.  The eigenvalues of the map $[\psi]$ 
 are all that is needed to express any function of the entanglement of the bipartite state $\ket{\psi}$.  
\end{remark}

\begin{remark}[Extending Lemma \ref{lemma:graphical-map-state}]
One can readily extend Lemma \ref{lemma:graphical-map-state} and consider instead the scenario given in the following Figure. 
 \begin{figure}[H]
 \begin{center}
  {\center  \includegraphics[width=12cm]{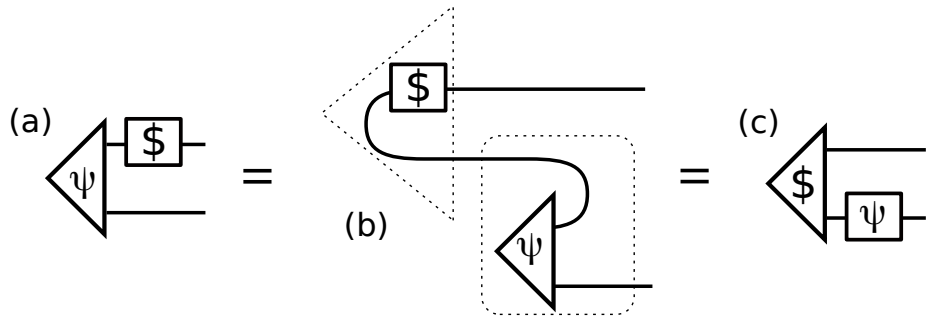}}  
 \end{center}
 \caption{Graphical formulation of channel state duality.  (a) The diagram representing $\left(\op{\$} \otimes \op{1}\right) \ket{\psi}$.  (b) The tensor network of the same state and the application of 
  the natural equivalence induced by the snake equation per Lemma \ref{lemma:graphical-map-state}. (c) We then think of the state $\psi$ as a map $\sum \psi_{ij}\ket{i}\bra{j}$ acting on the 
  state $\ket{\$}$. \label{fig:result}} 
 \end{figure}

\end{remark}

\section{Entanglement Evolution}\label{sec:evolution}

\begin{remark}[Background reading] 
For more details on standard results in the area of open systems theory, see for instance \cite{Bengtsson2006}.  For result results unifying several pictures of open systems evolutions see \cite{wood11}. 
\end{remark}

Quantum operations are maps acting on quantum states $\rho$ such that they preserve the positivity and boundedness of the corresponding density operator \cite{Bengtsson2006}. Such maps are known as \textit{completely positive}. They can be used to represent the action of unitary maps or, more generally, decoherence and measurement or combinations thereof. In general a completely positive map $\op{\$$}, called a \textit{superoperator}, may be described in terms of Kraus operators $\op{A}_k$,
\begin{align}
	\op{\$}\left[\rho\right] = \sum_k \op{A}_k \rho \op{A}_k^\dagger.
\end{align}
The map $\op{\$}$ is said to be trace-preserving iff $\Tr(\rho) = \Tr(\op{\$} \rho)$ for all positive operators $\rho$ and it can be shown that this is true iff $\sum_k \op{A}_k\op{A}_k^\dagger = \op{1}$. Such maps faithfully preserve the normalization of $\rho$. 

We consider bipartite states $\rho \in \spc{H}_A \otimes \spc{H}_B$. Since such states may be entangled, one is often particularly concerned with how entanglement evolves or changes under completely positive maps. Thus, given an entanglement measure $\2C$ and a \emph{local} completely positive map $\op{\$} = \op{\$}_A \otimes \op{\$}_B$ acting as $\op{\$}\left[\rho\right] = \sum_k (\op{A}_k\otimes \op{B}_k) \rho (\op{A}_k^\dagger\otimes \op{B}_k^\dagger)$ we will want to compute $\2C[(\op{\$}_A \otimes \op{\$}_B) \rho ]$, for some entanglement measure $\2C$. An elementary, but still critically important, example of such maps are one-sided maps of the form $\op{\$} = \op{\$}_A \otimes \op{1}_B$.

Entanglement evolution under one-sided maps has been characterised in the literature. In \cite{Konrad08} Konrad et al. showed that for pure bipartite qubit states 
\begin{align}
\2C\bigl[(\op{\$}_A \otimes \op{1}) \ket{\psi}\bra{\psi}\bigr] = \2C\left[(\$_A
\otimes \op{1}) \uket{\Phi^+}\ubra{\Phi^+}\right] \cdot 
\2C\bigl[\ket{\psi}\bra{\psi}\bigr]
\end{align} 
and so they factored the function $\2C[\$, \ket{\psi}]$ of two variable arguments into a product of functions, one depending only on the starting state and the other only on the completely positive map characterising the one-sided evolution. Thus, calculating the evolution of entanglement for a single initial state, $\ket{\Phi^+}$, allows us to compute it for all other pure initial states.  As will be seen later, this provides an upper bound on the quantity of evolved entanglement for all initial mixed states.  Here 
\begin{align} 
\ket{\Phi^+} = \frac{1}{\sqrt{2}}\left(\ket{00} + \ket{11}\right)
\end{align}
is again the typical Bell state and $\2C$ is the concurrence. For mixed states they found the upper bound to be given by the product \cite{Konrad08}
\begin{align}
\2C\bigl[(\$_A \otimes \op{1}) \rho \bigr] \leq \2C\left[(\$_A \otimes \op{1})
\uket{\Phi^+}\ubra{\Phi^+}\right] \cdot  \2C[\rho ].
\end{align} 


In \cite{Tiersch09} the analogous result was obtained for
higher-dimensional systems, but where $\2C$ was replaced by the $\2G$-concurrence,
defined for $d$-dimensional pure states 
\begin{align}
\ket{\psi} = \sum_{i,j = 1}^d
A_{ij} \ket{i} \otimes \ket{j}
\end{align}
with $A:=[\psi]$ as 
\begin{align}
\2C_d[\ket{\psi}] = d\cdot 
\left[\det(A^\dagger A) \right]^{1/d} 
\end{align}

For mixed states the $\2G$-concurrence is defined through the convex roof construction \cite{Gour05}
\begin{align} 
\2G_d[\rho] = \inf \sum_i p_i \2C_d\bigl[\ket{\psi_i}\bigr]
\end{align} 
where the infimum is taken over all convex decompositions of $\rho$ into pure states, i.e., all decompositions of the form
\begin{align} 
\rho = \sum_i p_i \ket{\psi_i}\bra{\psi_i}
\end{align} 
with $p_i > 0$, $\sum_i p_i =
1$ and the $\ket{\psi_i}$ need not be orthogonal. Hence, we note in particular that it is not sufficient to consider only the spectral decomposition. 

The results on factorisation of entanglement evolution were further extended to the multipartite case in \cite{Gour10}, where Gour showed that the same factorization result holds for all SL-invariant entanglement measures. These are defined as:
\begin{definition}[SL-invariant measures \cite{Gour05, Gour10}]\label{def:SLinvariance} 
Let $G$ be the group $G = \grp{SL}(d_1, \mathbb{C}) \otimes \grp{SL}(d_2, \mathbb{C}) \otimes \ldots \otimes \grp{SL}(d_n, \mathbb{C})$. The groups $\grp{SL}(d_k, \mathbb{C})$ are the special linear groups represented by all $d_k \times d_k$ matrices with determinant 1. A SL-invariant measure of multipartite entanglement $\2C: \spc{B}(\spc{H}) \rightarrow \mathbb{R}^+$ maps the bounded operators on the Hilbert space $\mathcal{H}$ (not necessarily normalised density matrices) to non-negative real numbers. $\2C$ must satisfy the following properties:
\begin{enumerate} 
	\item For every $g \in G$ and $\rho \in \spc{B}(\spc{H})$ we have that $\2C[g \rho g^\dagger] = \2C[\rho]$.
	\item It is homogenous of degree 1, $\2C[r \rho] = r \2C[\rho]$.
	\item For all valid mixed states it is given in terms of the convex roof construction: \begin{align} \2C[\rho] = \min \sum_i p_i \2C\bigl[\uket{\psi_i}\ubra{\psi_i}\bigr]. \end{align} Here the minimum is taken over all convex decompositions of $\rho$, i.e. all decompositions into $\rho = \sum_i p_i \ket{\psi_i} \bra{\psi_i}$ where $\braket{\psi_i}{\psi_j}$ is not necessarily zero. 
\end{enumerate}
\end{definition}

Now we will show how the above results, both those for mixed and pure states and generally for qudits, can be described through the use of Penrose graphical calculus and the Choi-Jamio{\l}kowski isomorphism. More precisely, we will show that
\begin{align}\label{eqn:product-structure}
	\2C[(\op{\$}_A \otimes \op{1}) \ket{\psi}\bra{\psi}] = \2C[(\$_A
\otimes \op{1}) \ket{\Phi^+}\bra{\Phi^+}] \cdot 
\2C[\ket{\psi}\bra{\psi}].
\end{align}
The factorised form of the entanglement evolution is desirable as it simplifies and unifies calculations of entanglement once a one-sided map has been applied. 
To arrive at the factorisation formula we first require one further result. 

Given an $n \times n$ matrix $\op{A}$ with $\det(\op{A}) \neq 0$ and an SL-invariant n-dimensional entanglement measure $\2C$, we find that 
\begin{equation}\label{eqn:FactorisationLemma}
\2C\bigl[\op{A} \rho \op{A}^\dagger] = \2C\bigl[|\det(\op{A})|^{2/n} g \rho g^\dagger\bigr] = \left|\det(\op{A})^{2/n}\right| \2C[\rho].
\end{equation}
where we have defined $\op{A} := \det(\op{A})^{1/n} g$, with $g \in \grp{SL}(n)$. 

 As will be seen, this calculation \eqref{eqn:FactorisationLemma} will be applied to maps $\op{A}$ that arise under wire bending duality from states.  Consider 
 \begin{center}
 \includegraphics[width=6cm]{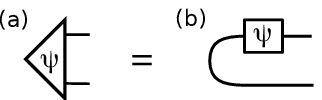}
\end{center}
 and by Lemma \ref{lemma:graphical-map-state} we establish equality of (a) and (b).  Hence, every bipartite quantum state $\ket{\psi}$ can be represented 
 as the generalized Bell state $\ket{\Phi^+}$ acted on by a single sided map $[\psi]$.  Let $\op{A}:=[\psi]$ then the rank of $\op{A}$ determines entanglement properties of $\ket{\psi}$.  $\op A$ is full rank iff $\ket{\psi}$ is SLOCC equivalent to $\ket{\Phi^+}$.
 In fact, $\op A$ is full rank iff $\det(\op{A})\neq 0$ or equivalently iff $\det(\op{A}^\dagger\op{A}) > 0$.  This will soon become relevant as we will use this duality to view 
 $\ket{\psi}$ as a single sided map and rely on the determinant to evaluate the entanglement measure of interest.

\begin{figure}[h]
\begin{center}
 {\center  \includegraphics[width=12cm]{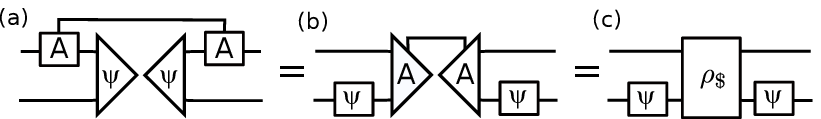}}  
\end{center}
\caption{Tensor network summary of the crucial duality behind the theory of entanglement evolution.  (a) State $\ket{\psi}$ acted on by a single sided superoperator $\$$. (b) Application of Lemma \ref{lemma:graphical-map-state} establishes a duality mapping the scenario in (a) into view where $\ket{\psi}$ becomes a single sided superoperator $[\psi]$ which acts on the bipartite mixed state $\rho_\$ = \sum_k \ket{\op{A}_k}\bra{\op{A}_k}$.  (c) An equivalent expression for the valence four tensor $\rho_\$$.  
\label{fig:summary}} 
\end{figure}

Now that we have established a factorisation of the above formula \eqref{eqn:FactorisationLemma} we will show that the factor $\det(\op{A})^{2/n}$ is nothing other than $\2C[\left(\$ \otimes \op{1}_B\right) \ket{\phi^+}\bra{\phi^+}]$. Let us then consider evolution of the bipartite state $\ket{\psi}$ under the superoperator $\$$, with a tensor network given as
\begin{center}
\includegraphics[width=4cm]{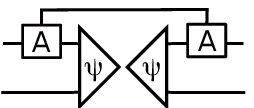}    
\end{center}
We then apply Lemma \ref{lemma:graphical-map-state} to both sides and arrive at 
\begin{center}
 \includegraphics[width=4cm]{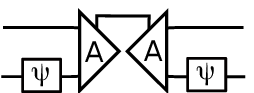}
\end{center}
So $\ket{\psi}$ becomes a superoperator acting on the density operator 
\begin{equation}
 \rho_\$ = \sum_k \ket{\op{A}_k}\bra{\op{A}_k} 
\end{equation}
which can be equivalently expressed in Figure \ref{fig:summary} (c) where 
the remaining subfigures (a, b) summarize the duality behind entanglement evolution.   
Hence, Lemma 
\ref{lemma:graphical-map-state} allows us to show the equivalence of the states (a) and (c) in Figure \ref{fig:result} and from this it 
follows that 
\begin{equation} 
\left(\op{\$} \otimes \op{1}\right) \ket{\psi}\bra{\psi} = \left(\op{1} \otimes
\op{\$}_{\psi}\right) \rho_{\$}. 
\end{equation}
which is alternatively expressed as $(\eye \otimes [\psi]) \sum_k \ket{\op{A}_k}\bra{\op{A}_k}$.  

Furthermore, notice that any state $\ket{\psi}$ may
be written using the Schmidt decomposition as $\ket{\psi} = \sum_k
\sqrt{\omega_k} \ket{kk}$. Thus, the map $\op{\$}_\psi$ may be considered simply as an action of a matrix
\begin{equation} 
\op{A} = \sqrt{n}\sum_k \sqrt{\omega_k} \ket{k}\bra{k},
\end{equation} 
where the factor of $\sqrt{n}$ enters because of normalization. Because of the purity
of $\ket{\psi}$ we arrive at only a single Kraus operator. Thus we find that
\begin{align}
	\2C[(\op{\$} \otimes \op{1}) \ket{\psi}\bra{\psi}] = \det(\op{A})^{2/n} \2C[\rho_\$ ]\label{eq:IntermedEvol}
\end{align}
Notice, that $\det(\op{A})^{2/n} = n \prod_{k=1}^n \omega_k^{1/n}$. This is exactly the definition of $\2G$-concurrence, reflecting the fact that $\2G$-concurrence is the only SL-invariant measure of entanglement for bipartite qudits. The second factor in the above equation (\ref{eq:IntermedEvol}) is simply equivalent to 
\begin{align}
	\2C[\rho_\$] = \2C\bigl[(\op{\$} \otimes \op{1})\uket{\Phi^+}\ubra{\Phi^+}\bigr]
\end{align} where
\begin{align}
	\ket{\Phi^+} = \frac{1}{\sqrt{n}}\sum_{k=1}^n \ket{kk}.
\end{align}
We can therefore rewrite the equation as
\begin{align}\label{eqn:ev}
	\2C[(\op{\$} \otimes \op{1}) \ket{\psi}\bra{\psi}] = \2C\bigl[(\op{\$} \otimes \op{1})\uket{\Phi^+}\ubra{\Phi^+}\bigr] \cdot \2C\bigl[\ket{\psi}\bra{\psi}\bigr].
\end{align}
\begin{equation}
 = \2C\bigl[(\op{\$} \otimes \op{1})\uket{\Phi^+}\ubra{\Phi^+}\bigr] \cdot n \prod_{k=1}^n \omega_k^{1/n}
\end{equation}

The term $\2C\bigl[(\op{\$}\otimes \op{1})\uket{\Phi^+}\ubra{\Phi^+}\bigr]$ still needs to be evaluated. However, its value tells us how concurrence evolves under $\op{\$}\otimes\op{1}$ for all pure states. It thus only needs to be computed once. Here the entanglement measure $\2C$ is now understood to be the $\2G$-concurrence. The equation factorises concurrence into a part entirely due to evolution dynamics and a part entirely due to the initial state. 

Having established the evolution equation \eqref{eqn:ev}, we will now consider an upper bound. Since the measure $\2C$ is given by a convex roof construction for mixed states, it is a convex measure --- i.e.
\begin{align}
	\2C\Bigl[\sum_k p_k \rho_k\Bigr] \leq \sum_k p_k \2C[\rho_k].
\end{align}
We can use the convexity of the measure $\2C$ to find an upper bound on the entanglement evolution for mixed states. This upper bound is given by
\begin{align}
	\2C \bigl[(\op{\$} \otimes \op{1}) \rho \bigr] \leq \2C\bigl[(\op{\$} \otimes \op{1})\uket{\Phi^+}\ubra{\Phi^+}\bigr] \cdot \2C[ \rho ].
\end{align}

%
%

\section{Conclusion}\label{sec:conclusion}


We have shown how entanglement evolution can be understood and even unified in terms of a tailored variant of the Penrose graphical calculus. We developed a key graphical tool to simplify the derivation of several important recent result in quantum information theory --- that of entanglement evolution of bipartite quantum states. The result was originally obtained through the Choi-Jamio{\l}kowski isomorphism, which we have shown in the diagrammatic language reduces to bending a contracting wire into a snake.  We introduced the necessary tensor networks and showed how the evolution of $\2G$-concurrence, the only SL-invariant measure of bipartite entanglement, can be obtained from these network diagrams.  We thus developed a tensor network approach that, through the use of Penrose duality, recovers the known theory of entanglement evolution. This goal necessitated the introduction of several new methods which could prove useful in other areas. In particular, we anticipate the existence of other problems that could be simplified through the use of the diagrammatic variant of channel state duality we have developed for our purposes here. 
This cross pollination could lead to further insights into the strengths and limitations of the approach we have presented here.   

As for the entanglement evolution itself, it is now fairly well understood for pure initial states and one-sided channels, few results are known for more general maps and for mixed initial states. Progress in this direction through the use of the Choi-Jamio{\l}kowski technique is inhibited by the fact that the dual equivalent of a mixed state is a super-operator rather than a simple matrix. Mathematically, given a mixed state in its spectral decomposition $\rho = \sum_k p_k \ket{\psi_k}\bra{\psi_k}$, its wire bent dual is the map (see part (a) of figure \ref{fig:summary})
\begin{align}
	\$_\rho = \sum_k p_k \op{A}_k \cdot \op{A}_k^\dagger,
\end{align}
where $\op{A}_k$ is the wire bent dual of the state $\ket{\psi_k}$. With pure states we have the fortune that $\$_\rho$ reduces to a single matrix and we can thus greatly simplify the calculation. The fact that it is now a genuine super-operator limits the applicability of this approach. One would thus expect that a qualitatively different approach is needed to find a closed form expression for entanglement evolution of mixed states. And in fact, the very existence of such an expression is not guaranteed. However, finding it would immediately also yield a closed form expression for the evolution under maps acting on more than one subsystem $\op{\$}_1 \otimes \op{\$}_2$. To see this, notice that $\op{\$}_1 \otimes \op{\$}_2 = \left(\op{\$}_1 \otimes \op{1}\right) \cdot \left(\op{1} \otimes \op{\$}_2\right)$ thus treating the map as a sequence of two maps acting on only one subsystem.

As with the purely algebraic approach, the drawback of the tensor network approach is that it relies on the simplifying properties of bipartite SL-invariant
measures of entanglement.  The key contribution being a conceptual aid and exposure of the underlying mathematical structure at play. Namely, the property that acting on a single subsystem of a quantum state with a non-singular matrix as $\left(\op{A} \otimes \op{1}\right) \rho \left(\op{A}^\dagger \otimes \op{1}\right)$ multiplies an SL-invariant entanglement measure with a function of $\det(A)$. Other entanglement measures may not satisfy this property, and indeed the usual concurrence only satisfies it in two dimensions. 

Acknowledgments.  We would like to thank Michele Allegra, Dieter Jaksch,  Stephen Clark, Mark M. Wilde, James Whitfield and Zoltan Zimboras.

\end{document}